\documentclass[aps,prd,groupedaddress,preprint,eqsecnum,nofootinbib]{revtex4}
\usepackage{graphicx,epsf,amssymb,amsbsy,amsfonts,amssymb,amsmath,physics,courier}
\usepackage{hyperref}
 \usepackage{xcolor}\hypersetup{linkbordercolor=blue}

\def\be{\begin{equation}}
\def\ee{\end{equation}}

\def\f{\phi}

\def\bg{\bar{g}}
\def\beq{\begin{eqnarray}}\def\eeq{\end{eqnarray}}
\def\ba#1\ea{\begin{align}#1\end{align}}
\def\bg#1\eg{\begin{gather}#1\end{gather}}
\def\bm#1\em{\begin{multline}#1\end{multline}}
\def\bmd#1\emd{\begin{multlined}#1\end{multlined}}

\def\D{\Delta}

\def\G{\Gamma}

\def\({\left(}
\def\){\right)}
\def\[{\left[}
\def\]{\right]}

\def\f{\frac}

\def\D{\Delta}

\pdfoutput=1

\begin{document}
\hfuzz 9pt
\title{An infinite family of $w_{1+\infty}$ invariant theories on the celestial sphere}
\author{Shamik Banerjee}
\affiliation{National Institute of Science Education and Research (NISER), Bhubaneswar 752050, Odisha, India \\ and \\ Homi Bhabha National Institute, Anushakti Nagar, Mumbai, India-400085}
\email{banerjeeshamik.phy@gmail.com}
\author{Harshal Kulkarni}
\affiliation{Department of Physical Sciences, IISER Kolkata, Mohanpur, West Bengal 741246, India}
\email{hck18ms056@iiserkol.ac.in}
\author{Partha Paul}
\affiliation{Centre for High Energy Physics, Indian Institute of Science,
C.V. Raman Avenue, Bangalore 560012, India.}
\email{pl.partha13@gmail.com}

\begin{abstract}
In this note we determine the graviton-graviton OPE and the null states in any $w_{1+\infty}$ symmetric theory on the celestial sphere. Our analysis shows that there exists a discrete \textit{infinite} family of such theories. The MHV-sector and the quantum self dual gravity are two members of this infinite family. Although the Bulk Lagrangian description of this family of theories is not currently known to us, the graviton scattering amplitudes in these theories are heavily constrained due to the existence of null states. Presumably they are exactly solvable in the same way as the minimal models of $2$-D CFT.  

\end{abstract}

\maketitle
\tableofcontents

\section{Introduction}
Celestial holography \cite{Strominger:2017zoo} is an attempt to formulate a dual theory of quantum gravity in asymptotically flat space time. When the bulk is $(3+1)$ dimensional, the dual theory is conjectured to be two dimensional and it lives on the celestial sphere. The correlation functions of this theory are given by the Mellin transform of the bulk $S$-matrix elements \cite{Pasterski:2016qvg,Pasterski:2017kqt,Banerjee:2018gce} and an important question is how to compute them in the dual formulation. 

Now in the absence of a Lagrangian description one can resort to the Bootstrap technique and the job is sometimes simplified if the theory has a large number of symmetries. For example, one can solve the minimal models in $2$-D CFT by using the representation theory of Virasoro and other current algebras even when the Lagrangian description is not known. In celestial holography the two dimensional dual theory also has various (infinite dimensional) current algebra symmetries \cite{Sachs:1962zza,Strominger:2013jfa,He,Barnich:2009se,Barnich:2011ct,Kapec:2016jld,Kapec:2014opa,He:2017fsb,Banerjee:2020zlg,Guevara:2021abz,Stieberger:2018onx,Banerjee:2022wht,Banerjee:2021cly,Strominger:2021lvk,Donnay:2020guq,Himwich:2021dau,Ball:2021tmb,Adamo:2021lrv,Costello:2022wso,Gupta:2021cwo,Freidel:2021dfs,Freidel:2021ytz} and it is expected that the bootstrap approach should simplify considerably in this case\footnote{Please see \cite{Atanasov:2021cje} for a discussion of conformal bootstrap approach to celestial holography. Also recently celestial holography has been used to shed light on $S$-matrix bootstrap in \cite{Ghosh:2022net}.}. 

In this paper we consider dual theories which are invariant under the $w_{1+\infty}$ symmetry\footnote{To be more precise the wedge subalgebra of $w_{1+\infty}$.}\cite{Guevara:2021abz,Strominger:2021lvk,Himwich:2021dau,Ball:2021tmb,Adamo:2021lrv,Costello:2022wso}. We determine the graviton-graviton OPE and the null states in such theories. So far two examples of such theories were known. One of them is the MHV-sector \cite{Banerjee:2020zlg,Banerjee:2021cly} and the other one is the quantum self dual gravity \cite{Ball:2021tmb}. However, we find that there is a \textit{discrete infinite family} of such theories and the MHV-sector and the quantum self dual gravity are two members of this family. Bulk description of this family of theories is not known to us and we leave this to future work. 

\section{General structure of $w$-invariant OPE}

The general structure of the $w$-invariant OPE between two positive helicity outgoing graviton primaries with weights $\D_1$ and $\D_2$ can be written as,
\be\label{uni}
\begin{gathered}
G^{+}_{\D_1}(z_1,\bar z_1) G^+_{\D_2}(z_2,\bar z_2) = -\frac{\bar z_{12}}{z_{12}} \sum_{n=0}^{\infty} B\(\D_1 -1 +n, \D_2 -1\) \frac{{\bar z_{12}}^n}{n!} \bar\partial^n G^+_{\D_1+\D_2}(z_2,\bar z_2) \\ 
+ \sum_{p,q=0}^{\infty} z_{12}^p\bar z_{12}^q \sum_{k=1}^{\tilde n_{p,q}} \tilde C^k_{p,q} (\D_1,\D_2) \tilde O^{p,q}_k(\D_1,\D_2,z_2,\bar z_2)
\end{gathered}
\ee
where the first line contains the universal terms \cite{Guevara:2021abz} having simple pole in $z_{12}$. The second line contains the non-singular terms in $z_{12}$ and $\bar z_{12}$ and they are linear combinations of the $w$ descendants of a positive helicity graviton primary, denoted by $\tilde O^{p,q}_i$. The positive integer $\tilde n_{p,q}$ may be finite or infinite. 
Our goal is to determine the operators $\tilde O^{p,q}_i$ and the OPE coefficients $\tilde C^i_{p,q}(\D_1,\D_2)$ which can appear in a $w$-invariant theory. Now before we proceed let us briefly describe the symmetry algebra that follows from the universal singular terms of the OPE \eqref{uni}. 

So we start by defining the tower of conformally soft \cite{Donnay:2018neh} gravitons \cite{Guevara:2021abz}
\be
H^k(z,\bar z) = \lim_{\D \rightarrow k} (\D-k) G^+_{\D}(z,\bar z), \ k =1,0,-1,-2,...
\ee
with weights $\(\frac{k+2}{2}, \frac{k-2}{2}\)$. It follows from the structure of the singular terms in \eqref{uni} that we can introduce the following truncated mode expansion
\be
H^k(z,\bar z) = \sum_{m = \frac{k-2}{2}}^{\frac{2-k}{2}} \frac{H^k_m(z)}{\bar z^{m + \frac{k-2}{2}}}
\ee
and the modes $H^k_m(z)$ are the conserved holomorphic currents. The currents $H^k_m(z)$ can be further mode expanded 
\be
H^k_m(z) = \sum_{\alpha \in \mathbb{Z} - \frac{k+2}{2}} \frac{H^k_{\alpha,m}}{z^{\alpha+ \frac{k+2}{2}}}
\ee
and one can show \cite{Guevara:2021abz} that the modes $H^k_{\alpha,m}$ satisfy the algebra\footnote{Here we are assuming that $\kappa=\sqrt{32\pi G_N}=2$.}
\be\label{hsa}
\begin{gathered}
\[ H^k_{\alpha,m}, H^l_{\beta,n}\] \\ = - \[ n(2-k) - m(2-l)\] \frac{\( \frac{2-k}{2} -m + \frac{2-l}{2} - n -1\)!}{\( \frac{2-k}{2} -m \)! \( \frac{2-l}{2} - n\)!}\frac{\( \frac{2-k}{2} +m + \frac{2-l}{2} +n -1\)!}{\( \frac{2-k}{2} +m \)! \( \frac{2-l}{2} +n\)!} H^{k+l}_{\alpha+\beta, m+n}
\end{gathered}
\ee
This is called the Holographic Symmetry Algebra (HSA). Now if we make the following redefinition (or discrete light transformation)\cite{Strominger:2021lvk}
\be
w^p_{\alpha, m} = \frac{1}{2}\(p-m-1\)! \(p+m-1\)! H^{-2p+4}_{\alpha,m}
\ee
then \eqref{hsa} turns into the $w_{1+\infty}$ algebra\footnote{This is the wedge subalgebra of $w_{1+\infty}$.}
\be\label{w}
\[w^p_{\alpha,m}, w^q_{\beta,n}\] = \[ m(q-1) - n(p-1)\] w^{p+q-2}_{\alpha+\beta,m+n}
\ee
For our purpose it is more convenient to work with the HSA \eqref{hsa} rather than the $w_{1+\infty}$ algebra. However, we continue to refer to the HSA as the $w$ algebra. 

Now suppose we have two different $w$ invariant theories which we call $A$ and $B$. $A$ and $B$ have \textit{identical} symmetry algebras \eqref{hsa} because there are no central charges and moreover, their operator contents are also the same. Therefore if the OPE is completely determined by $w$ covariance then there must exist a \textit{universal form} of the OPE which holds in \textit{both} theory $A$ and theory $B$. But how is this possible if the graviton scattering amplitudes in theory $A$ and theory $B$ are different ? This is still possible if the following holds: 

Let us denote the universal OPE by $\text{OPE}_{w}$ and the OPEs obtained from the graviton scattering amplitudes of theory A and theory B by $\text{OPE}_A$ and $\text{OPE}_B$, respectively. Note that $\text{OPE}_{w}$ holds in both theory $A$ and theory $B$. Therefore it must be true that
\be\label{theory}
\begin{gathered}
\text{OPE}_{w} = \text{OPE}_A + \text{Linear combination of Null-States of theory A} \\
\text{OPE}_{w} = \text{OPE}_B + \text{Linear combination of Null-States of theory B}
\end{gathered}
\ee
This is a major simplification if we know the OPE and null states of either theory $A$ or theory $B$. 

\subsection{What is theory $A$?}
Our task will be simplified if we can find a theory $A$ such that the null states of theory $B$ can be written as a linear combination of the null states of theory $A$. This does not mean that the theory $A$ and theory $B$ have the same null states but, this implies that the null states of theory $A$ form a basis in terms of which the null states of theory $B$ can be expanded. Once we find such a theory $A$ we can write, as a consequence of \eqref{theory},
\be
\boxed{
\text{OPE}_{B} = \text{OPE}_A + \text{Linear combination of Null-States of theory A}}
\ee

Now in our case we take the theory $A$ to be the MHV-sector. This is the simplest theory in the sense that although this is $w$ invariant, the graviton-graviton OPE can be written as \cite{Banerjee:2020zlg} a linear combination of supertranslation and $\overline{sl_2}$ current algebra\footnote{Supertranslation and $\overline{sl_2}$ current algebras are subalgebras of the $w_{1+\infty}$ algebra. They are generated by the soft gravitons $H^1(z,\bar z)$ and $H^0(z,\bar z)$.} descendants only. Therefore MHV sector has the maximum number of null states, i.e, maximally constrained among all the $w$ invariant theories. Now the OPE and the null states in the MHV-sector are known \cite{Banerjee:2020zlg,Banerjee:2021cly} and we will use them to construct the general $w$ invariant OPE. 

\section{Null states in the MHV sector}
According to our discussion in the previous section, we can write
\be
\label{wope}
\begin{gathered}
G^{+}_{\D_1}(z_1,\bar z_1) G^+_{\D_2}(z_2,\bar z_2) |_B = G^{+}_{\D_1}(z_1,\bar z_1) G^+_{\D_2}(z_2,\bar z_2) |_{\text{MHV}} \\ 
+ \sum_{p,q=0}^{\infty} z_{12}^p\bar z_{12}^q \sum_{k=1}^{n_{p,q}} C^k_{p,q} (\D_1,\D_2) N^{p,q}_k(z_2,\bar z_2)
\end{gathered}
\ee
where $N^{p,q}_k$ are the null states of the MHV sector which can appear at $\mathcal{O}(z^p\bar z^q)$ and the integers $n_{p,q}$ can be finite or infnite. Since the MHV-sector OPE contains the universal singular part of $\text{OPE}_B$, the second line of \eqref{wope} contains only the non-singular terms.  

We now discuss the MHV null states at $\mathcal{O}(z_{12}^0\bar z_{12}^0)$ and at $\mathcal{O}(z_{12}^0 \bar z_{12}^1)$. This can be done at any order however, for the sake of simplicity, we focus on these two orders only. Moreover, the $\mathcal{O}(z_{12}^0\bar z_{12}^1)$ terms in the $\text{OPE}_B$  give rise to interesting constraints on the graviton scattering amplitudes \cite{Banerjee:2020zlg} in theory $B$. So it will be interesting to determine them.

\subsection{$N^{0,0}_k$}
The MHV null states that can appear at $\mathcal{O}({z_{12}^0\bar z_{12}^0})$ are given by \cite{Banerjee:2020zlg,Banerjee:2021cly}\footnote{The null states are obtained by starting from the graviton-graviton OPE in the MHV sector and then making one of the gravitons conformally soft. This is discussed in detail in \cite{Banerjee:2020zlg,Banerjee:2021cly}.},
\be
\Phi_k(\D) = \[ H^{1-k}_{\f{k-3}{2},\f{k+1}{2}} \(-H^{1}_{-\f{1}{2},-\f{1}{2}}\)^k -\f{(-1)^k}{k!} \f{\G(\D +k-2)}{\G(\D-2)}H^{1}_{-\f{3}{2},\f{1}{2}} \] G^+_{\D-1}
\ee
where $k=1,2,3,\cdots, \infty$. However, it is more convenient to work with the new basis defined by 
\be
\begin{gathered}
\Omega_k(\D) = \sum_{n=1}^{k} \f{1}{(k -n)!} \f{\G(\D+k-2)}{\G(\D +n-2)}\Phi_n(\D)
\end{gathered} 
\ee
The physical significance of this basis will become clear in the next section. 

\subsection{$N^{0,1}_k$}
The MHV null states that can appear at $\mathcal{O}({z_{12}^0\bar z_{12}^1})$ are given by \cite{Banerjee:2020zlg,Banerjee:2021cly}, 
\be
\begin{gathered}
\Psi_k(\D) = \bigg[H^{-k}_{\f{k-2}{2},\f{k}{2}}\(-H^{1}_{-\f{1}{2},-\f{1}{2}}\)^{k+1} - \f{(-1)^k}{k!}\f{\G(\D+k-2)}{\G(\D-2)} H^{0}_{-1,0}\(-H^{1}_{-\f{1}{2},-\f{1}{2}}\)\\ - \f{(-1)^k k}{(k+1)!} \f{\G(\D+k-2)}{\G(\D-3)} H^{1}_{-\f{3}{2},-\f{1}{2}}\bigg] G^+_{\D-2} 
\end{gathered}
\ee
where $k=1,2,3 \cdots, \infty$. 
Again we introduce the new basis
\be
\Pi_k(\D) = \sum_{n=1}^{k} \f{1}{(k-n)!}\f{\G(\D+k-2)}{\G(\D+n-2)}\Psi_n(\D)
\ee
which is more convenient for our purpose.

There is a second set of null states at this order which involve the descendant $L_{-1}G^{+}_\D$. These are the null states of the Knizhnik-Zamolodchikov(KZ) type and their decoupling gives rise to differential equations for the scattering amplitudes \cite{Banerjee:2020zlg,Banerjee:2020vnt} which are the generalization of the standard KZ equation to Celestial holography. However, we do not need the KZ-type null states to write down the $\text{OPE}_B$ because the generator $L_{-1}$\footnote{$\bar L_{-1}$ is a generator of the $w$ algebra.} does not belong to the $w$ algebra. We will give the explicit form of these null states at a later part of the paper.

\section{Organization of the OPE at every order}\label{sl2v}

We have argued that the OPE in the theory $B$ can be written as,
\be
\begin{gathered}
G^{+}_{\D_1}(z,\bar z) G^+_{\D_2}(0,0) |_B = G^{+}_{\D_1}(z_1,\bar z_1) G^+_{\D_2}(0,0) |_{\text{MHV}} \\ 
+ \sum_{p,q=0}^{\infty} z^p\bar z^q \sum_{k=1}^{n_{p,q}} C^k_{p,q} (\D_1,\D_2) N^{p,q}_k(0,0)
\end{gathered}
\ee
where $N^{p,q}_k$ are the MHV null states. Now, an infinite number of null states can potentially appear at every order of the $\text{OPE}_B$. So we need some guiding principle which tells us which null states can appear at a given order and hopefully, we have only a finite number of them.  

The guiding principle is nothing but the $w$-covariance of the OPE. In fact, if we want to classify the states which can appear at a given order of the $\text{OPE}_B$, then we can work with the $sl_2(R)$ subalgebra of the $w$ algebra generated by the operators $\{H^1_{-1/2,-1/2}, \ H^0_{0,0}, \ H^{-1}_{1/2,1/2}\}$, 
\be\label{vsl}
\begin{gathered}
\[ H^0_{0,0}, H^1_{-1/2,-1/2}\] = H^1_{-1/2,-1/2} \\
\[ H^0_{0,0}, H^{-1}_{1/2,1/2}\] = - H^{-1}_{1/2,1/2} \\
\[H^1_{-1/2,-1/2}, H^{-1}_{1/2,1/2}\] = - H^0_{0,0}
\end{gathered}
\ee
We denote this by $sl_2(R)_V$\footnote{Here V stands for vertical. Please see Fig.\ref{fig} for an explanation. }.

Now using the relations 
\be
\begin{gathered}
H^0_{0,0} G^{+}_{\D}(z,\bar z) = 2\bar h G^{+}_{\D}(z,\bar z)= \(\D-2\) G^{+}_{\D}(z,\bar z) \\
H^1_{-1/2,-1/2} G^{+}_{\D}(z,\bar z) = - G^{+}_{\D+1}(z,\bar z) \\
H^{-1}_{1/2,1/2} G^{+}_{\D}(z,\bar z) = -\frac{1}{2} \hspace{0.03cm} \left[ (\D-2)(\D-3) + 4(\D-2)\bar z\bar{\partial} + 3\Bar{z}^{2}\Bar{\partial}^{2} \right] G^+_{\D-1}(z,\Bar{z})
\end{gathered}
\ee
and the $w$-covariance of the OPE, i.e, the fact that both sides of the OPE transform in the same way under $w$ transformations, one can easily check that:

 \textit{The MHV null states that can appear at $\mathcal{O}(z^p\bar z^q)$ of the OPE, transform in a representation of the $sl_2(R)_V$ subalgebra}. Moreover, this is the \textit{unique} $sl_2$ subalgebra with this property. 
 
We now verify this explicitly. 

\subsection{Action of $sl_2(R)_V$ on the null states}

\subsubsection{$N^{0,0}_k$}
Let us consider the action of the $sl_2(R)_V$ on the MHV null states $\Omega_k(\D)$ that can appear at $\mathcal{O}(z^0\bar z^0)$. It is given by, 
\be\label{f}
H^{1}_{-\f{1}{2},-\f{1}{2}}\Omega_k(\D) = - \Omega_k(\D+1)
\ee
\be\label{s}
\begin{gathered}
H^{-1}_{\f{1}{2},\f{1}{2}}\Omega_k(\D)=\f{1}{2}(k+1)(k+2)\Omega_k(\D-1) - \f{1}{2}(\D-4)(\D-5)\Omega_k(\D-1) \\ - \f{1}{2}(k+1)(k+2)\Omega_{k+1}(\D-1)
\end{gathered} 
\ee
and $H^0_{0,0} = 2 \bar L_0$ is obviously diagonal on these states. So we can see that the null states $\Omega_k(\D)$ indeed form a representation of $sl_2(R)_V$. However, this representation is reducible. For let us consider the subspace spanned by the (infinite) set of states $\{\Omega_{n+1}(\D),\Omega_{n+2}(\D),\Omega_{n+3}(\D), \cdots\}$ where $n\ge 0$. It is easy to see from \eqref{f} and \eqref{s} that this set is closed under the action of the $sl_2(R)_V$. Therefore the representation carried by the states $\{\Omega_k(\D)\}_{k=1,2,...}$ is reducible. As a consequence we can set the states 
\be\label{reduce}
\Omega_{k+1}(\D) = 0, \  k\ge n \ge 0
\ee
without violating the $sl_2(R)_V$ symmetry. After doing this we are left with $n$ null states $\{\Omega_1(\D),...,\Omega_n(\D)\}$ which, as a consequence of \eqref{s} and \eqref{reduce}, again form a \textit{representation} of $sl_2(R)_V$. The crucial point is that the value of the integer $n$ is \textit{not} fixed by the $sl_2(R)_V$ symmetry. 

Now for a given $n$, there is another set of $\lceil{\frac{n}{2}}\rceil$\footnote{$\lceil{\frac{n}{2}}\rceil = \text{Smallest integer}$ $\ge \frac{n}{2}$.} nontrivial\footnote{There are of course the $n$ states $\{\Omega_1(\D),...,\Omega_n(\D)\}$ which transform in a representation of $sl_2(R)_V$ but, we cannot set them to zero because that will lead us again to the MHV sector.} states $\{\chi^1_n(\D),...,\chi_n^{\lceil{n/2}\rceil}(\D)\}$ defined as
\be\label{add1}
\begin{gathered}
\chi^1_n(\D) = \sum_{p=1}^n \Omega_p(\D)\\
\chi^i_n(\D) = \sum_{p=i}^n\prod_{q=i}^{2i-2}(p-q) \Omega_p(\D), \  i = 2,3,...,\lceil{\frac{n}{2}}\rceil
\end{gathered}
\ee
which transform in a representation of the $sl_2(R)_V$ as a consequence of \eqref{reduce}. We can also set these states to zero 
\be\label{add2}
\chi^i_n(\D) = 0 
\ee
without violating the $sl_2(R)_V$ symmetry. 


\subsubsection{$N^{0,1}_k$}
The action of the $sl_2(R)_V$ on the MHV null states $\Pi_k(\D)$ which can appear at $\mathcal{O}(z^0\bar z^1)$ is given by, 
\be
H^{1}_{-\f{1}{2},-\f{1}{2}} \Pi_k(\D) = - \Pi_{k+1}(\D+1) 
\ee
and 
\be
\begin{gathered}
H^{-1}_{\f{1}{2},\f{1}{2}} \Pi_k(\D)=\f{1}{2}k(k+1)\Pi_k(\D-1) -\f{1}{2}(\D-4)(\D-5)\Pi_k(\D-1) \\ -\f{1}{2}(k-1)(k+2)\Pi_{k+1}(\D-1)
\end{gathered} 
\ee
Therefore we can see the states $\Pi_k(\D)$ form a representation of the $sl_2(R)_V$. This representation is reducible because the subspace spanned by the states $\{\Pi_{n+1}(\D),\Pi_{n+2}(\D),\Pi_{n+3}(\D),\cdots\}$ form a representation of the $sl_2(R)_V$. We can get a smaller representation spanned by the states $\{\Pi_1(\D),\cdots,\Pi_{n}(\D)\}$ if we set 
\be
\Pi_{k+1}(\D)=0, \  k\ge n\ge 0
\ee
There is no other nontrivial representation of $sl_2(R)_V$ in the subspace spanned by the states $\{\Pi_1(\D),...,\Pi_n(\D)\}$.

\section{w invariance}
So far we have discussed the invariance of the OPE under $sl_2(R)_V$. There is another $sl_2(R)$ subalgebra which is generated by the global (Lorentz) conformal transformations $\{ H^0_{0,1},H^0_{0,0},H^0_{0,-1}\}$. We call this $\overline{sl_2(R)}$ because this acts only on the $\bar z$ coordinate. Now the $w$ symmetry is generated by the infinite number of soft currents $\{H^k_{p}(z)\}$ where $k=1,0,-1,-2,...$ is the dimension $(\D)$ of the soft operator and $\frac{k-2}{2}\le p\le - \frac{k-2}{2}$. For a fixed $k$, the soft currents $\{H^k_{-\frac{k-2}{2}}(z),...,H^k_{\frac{k-2}{2}}(z)\}$ transform in a spin $\frac{2-k}{2}$ representation of the $\overline{sl_2(R)}$. 
\begin{figure}[h!]
\includegraphics[scale=0.6]{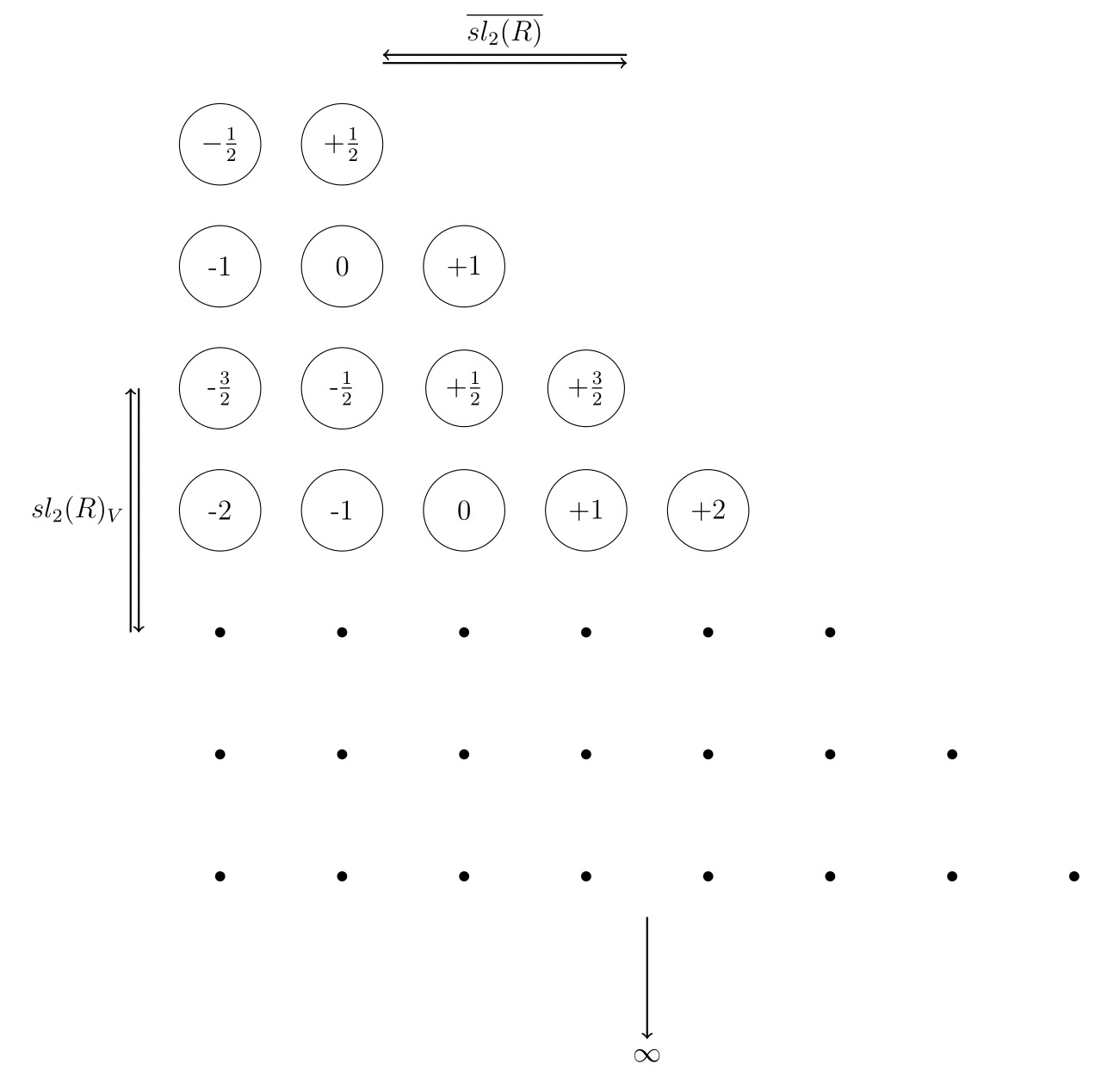}
\caption{The figure shows the soft currents. The rows and the columns are indexed by the $\overline{sl_2(R)}$ weights and the dimension $(\D=k=1,0,-1,-2,...)$ of the conformally soft graviton $H^k(z,\bar z)$, which generates the currents sitting in a row, respectively. $\overline{sl_2(R)}$ acts horizontally along a row and $sl_2(R)_V$ acts vertically along a column. In this way they generate the whole symmetry algebra starting from the current $H^1_{\frac{1}{2}}(z)$ on the top left corner.}
\label{fig}
\end{figure}
Now let us consider the currents $\{H^1_{\frac{1}{2}}, H^0_{1},...,H^k_{\frac{2-k}{2}},...\}$ with the \textit{lowest} $\overline{sl_2(R)}$ weights. These currents transform in an irreducible highest weight representation of the $sl_2(R)_V$. This can be seen from the following commutation relations following from \eqref{hsa},
\be
\begin{gathered}
\[ H^{-1}_{\frac{1}{2},\frac{1}{2}}, H^k_{\alpha,\frac{2-k}{2}}\] = - \frac{1}{2}(k-2)(k-3) H^{k-1}_{\alpha+\frac{1}{2},\frac{2-(k-1)}{2}}\\
\[ H^{0}_{0,0}, H^k_{\alpha,\frac{2-k}{2}}\] = (k-2) H^{k}_{\alpha,\frac{2-k}{2}}\\
\[ H^{1}_{-\frac{1}{2},-\frac{1}{2}}, H^k_{\frac{2-k}{2}}\] = - H^{k+1}_{\alpha-\frac{1}{2},\frac{2-(k+1)}{2}}
\end{gathered}
\ee
Therefore, starting from the current $H^1_{\frac{1}{2}}(z)$ we can generate \textit{any} other $w$ current by the combined action of the $\overline{sl_2(R)}$ and $sl_2(R)_V$ (Fig.\ref{fig}). 

Now, the way it is reflected in the structure of the OPE is the following. Suppose we consider the OPE of two positive helicity gravitons. The OPE is $w$ invariant if the singular terms have the following universal structure 
\be
G^{+}_{\D_1}(z,\bar z) G^+_{\D_2}(0,0) \supset -\frac{\bar z}{z} \sum_{n=0}^{\infty} B\(\D_1 -1 +n, \D_2 -1\) \frac{{\bar z}^n}{n!} \bar\partial^n G^+_{\D_1+\D_2}(0,0)
\ee
Now it has been shown in \cite{Pate:2019lpp} that the leading term in $\bar z$ is \textit{uniquely} determined by the $sl_2(R)_V$ invariance\footnote{It is assumed that the leading term is of $\mathcal{O}(\frac{\bar z}{z})$. This follows from the structure of the universal leading term in the collinear limit of two positive helicity gravitons.}. Once we know the leading term, the subleading terms in $\bar z$ of $\mathcal{O}(\frac{\bar z^q}{z}), q\ge 2$ are determined by the $\overline{sl_2(R)}$ invariance\footnote{Here the assumption is that the $\overline{\text{Vir}}$ is not part of the symmetry algebra.This is further discussed in \cite{Banerjee:2022wht}.}. Therefore if we make sure that the $sl_2(R)_V$ and the $\overline{sl_2(R)}$ symmetries are not broken then we are bound to have $w$ invariance. We now discuss the consequences of this. 

\subsection{$w$ invariance at $\mathcal{O}(z^0\bar z^0)$}
In section-\eqref{sl2v} we have shown that the equations
\be\label{omega1}
\Omega_{k+1}(\D) = 0, \  k\ge n \ge 0
\ee
are $sl_2(R)_V$ invariant. Now one can easily check that 
\be
H^0_{0,1} \Omega_k(\D) = 0 
\ee
where $H^0_{0,1}$ is, upto normalization, $\bar L_1$. So the MHV null states $\Omega_k(\D)$ are $\overline{sl_2(R)}$ primaries and as a result the equations \eqref{omega1} are $\overline{sl_2(R)}$ invariant. This implies that \eqref{omega1} is \textit{also} $w$ invariant.  

For the same reason, the additional null state relations \eqref{add2}
\be
\chi^i_n(\D) = 0, \  i= 1,2,...,\lceil{\frac{n}{2}}\rceil
\ee
at $\mathcal{O}(z^0\bar z^0)$ are also $w$ invariant. 

\subsection{$w$ invariance at $\mathcal{O}(z^0\bar z^1)$}
Similarly, we have shown in section-\eqref{sl2v} that the equations
\be\label{pi1}
\Pi_{k+1}(\D) = 0, \  k\ge n' \ge 0
\ee
are $sl_2(R)_V$ invariant. Now the action of $H^0_{0,1}$ on $\Pi_k(\D)$ is given by,
\be
\begin{gathered}
H^{0}_{0,1} \Pi_{k+1}(\D) = - (\D+k-2)\Omega_{k+1}(\D-1) + (k+3) \Omega_{k+2}(\D-1)
\end{gathered} 
\ee
So if we want to impose \eqref{pi1} without breaking the $\overline{sl_2(R)}$ invariance then we also have to impose
\be\label{omega2}
\Omega_{k+1}(\D) = 0, \  k\ge n' \ge 0
\ee
Therefore, \eqref{pi1} is $w$ invariant if \eqref{omega2} holds and we know from our previous analysis that \eqref{omega2} is indeed $w$ invariant. This also shows that $n_{0,0}=n_{0,1}$ in \eqref{wope} as a consequence of $w$ invariance.

\section{Finiteness and an infinite family of theories}
Our analysis in the last section shows that $w$ symmetry remains \textit{unbroken} if we set
\be
\Omega_{k+1}(\D) = 0, \  k\ge n \ge 0
\ee
and
\be
\Pi_{k+1}(\D) = 0, \  k\ge n \ge 0
\ee
\textit{simultaneously}. Therefore we can construct a $w$ invariant OPE at $\mathcal{O}(z^0\bar z^0)$ and $\mathcal{O}(z^0\bar z^1)$ by keeping only the \textit{finite}\footnote{There are still an infinite number of states $\{\Omega_p(\D+\mathbb{Z})\}_{1\le p\le n}$ required by the global time translation and the global subsubleading ($H^{-1}_{\frac{1}{2},\frac{1}{2}}$) invariance.} number of MHV null states $\{\Omega_1(\D),...,\Omega_n(\D)\}$ and $\{\Pi_1(\D),\cdots,\Pi_{n}(\D)\}$, respectively. The crucial point here is that the value of the integer $\textbf{n}$, is \textbf{not} fixed by $w$ invariance. So we get \textit{different} $w$ \textit{invariant} OPEs for different \textit{choices} of the integer $n$. For example, $n=0$ gives the MHV-sector. Similarly, direct calculation of graviton-graviton OPE using the scattering amplitudes shows that $n=4$ gives the OPE of the quantum self-dual gravity theory which is known to be $w$ invariant \cite{Ball:2021tmb}. The existence of this discrete infinite family of $w$ invariant OPEs is the main outcome of our analysis.

\section{$w$ invariant OPE coefficients}
In this section we write down the $w$ invariant OPE for a given value of $n$, at $\mathcal{O}(z^0\bar z^0)$ and $\mathcal{O}(z^0\bar z^1)$, using the states $\{\Omega_1(\D),...,\Omega_n(\D)\}$ and $\{\Pi_1(\D),\cdots,\Pi_{n}(\D)\}$. At $\mathcal{O}(z^0\bar z^0)$ we have
\be\label{0}
\begin{gathered}
G^{+}_{\D_1}(z,\bar z)G^+_{\D_2}(0,0)\big|_{\mathcal{O}(z^0\bar z^0)} = G^{+}_{\D_1}(z,\bar z)G^+_{\D_2}(0,0)\big|_{\text{MHV at} \ \mathcal{O}(z^0\bar z^0)} \\
+ \sum_{p=1}^n B(\D_1-1 + p,\D_2-1) \  \Omega_{p}(\D_1+\D_2)
\end{gathered} 
\ee
and at $\mathcal{O}(z^0\bar z^1)$
\be\label{1}
\begin{gathered}
G^+_{\D_1}(z, \bar z)G^+_{\D_2}(0,0)\big|_{\mathcal{O}(z^0\bar z^1)} = G^{+}_{\D_1}(z,\bar z)G^+_{\D_2}(0,0)\big|_{\text{MHV at} \  \mathcal{O}(z^0\bar z^1)} \\
+ \bar z \sum_{p=1}^n B(\D_1+p,\D_2-1) \ \Pi_p(\D_1+\D_2+1)
\end{gathered} 
\ee
The MHV sector OPE at $\mathcal{O}(z^0\bar z^0)$ and $\mathcal{O}(z^0\bar z^1)$ are given in \cite{Banerjee:2020zlg}.

 As we have already discussed the states $\{\Omega_1(\D),...,\Omega_n(\D)\}$ are not all independent because of the existence of the additional null state relations \eqref{add2}
\be
\chi^i_n(\D) = 0, \  i =1,2,...,\lceil{\frac{n}{2}}\rceil
\ee
We can further simplify \eqref{0} by using these null state relations. However, we choose to leave it in this form because the OPE coefficients are particularly nice when written in terms of these linearly dependent states. Now different values of $n$ give OPEs in different $w$ invariant theories.  We now discuss the KZ type null states of these theories. 

\section{Knizhnik-Zamolodchikov type null states}
KZ type null states occur at $\mathcal{O}(z^0\bar z^1)$ of the OPE. The simplest way to derive it is to demand the commutativity of the OPE, i.e, 
\be\label{comm}
G^+_{\D_1}(z_1, \bar z_1)G^+_{\D_2}(z_2,\bar z_2) = G^+_{\D_2}(z_2, \bar z_2)G^+_{\D_1}(z_1,\bar z_1)
\ee
If we use the consistency condition \eqref{comm} and the OPE given in \eqref{0} and \eqref{1} we get the following KZ type null state
\be\label{KZ}
\boxed{\Xi_n(\D) = \xi(\D) + \sum_{k=1}^n \Pi_k(\D+1) = 0}
\ee
where $\xi(\D)$ is the KZ type null state in the MHV sector given by \cite{Banerjee:2020zlg}
\be
\xi(\D) = \boxed{L_{-1}} G^{+}_{\D} + H^{0}_{0,-1}H^{1}_{-\f{3}{2},\f{1}{2}}G^+_{\D-1} + H^{0}_{-1,0}G^+_{\D} + (\D-1) \, H^{1}_{-\f{3}{2},-\f{1}{2}}G^+_{\D-1}
\ee
We have used that $\chi^1_n(\D)$ is a null state in this theory to arrive at the form \eqref{KZ}.

As consistency checks we can compute the actions of $\overline{sl_2(R)}$ and $sl_2(R)_V$ generators on $\Xi_n(\D)$. For example, 
\be
H^{0}_{0,1}\Xi_n(\D) = (n+2)\Omega_{n+1}(\D) - (\D-3) \chi^1_n(\D) = 0
\ee
because $\Omega_{n+1}(\D)$ and $\chi^1_n(\D)$ are both null states in this theory. Therefore $\Xi_n(\D)$ is a $\overline{sl_2(R)}$ primary. 

Similarly, we have 
\be
\begin{gathered}
H^{-1}_{\f{1}{2},\f{1}{2}}\Xi_n(\D) = -\f{1}{2}(\D-2)(\D-3)\Xi_n(\D-1) \\
 - \f{1}{2} (n+2)(n-1)\Pi_{n+1}(\D) - H^{-1}_{\f{1}{2},-\f{1}{2}} \chi^1_n(\D) - H^{0}_{0,-1} \( (\D-1)\chi^1_n(\D-1) + \chi^2_n(\D-1)\) \\
 \end{gathered} 
\ee
 But, since $\Pi_{n+1}(\D), \chi^1_n(\Delta)$ and $\chi^2_n(\D)$ are null states in the theory, we get 
\be
H^{-1}_{\f{1}{2},\f{1}{2}}\Xi_n(\D) = -\f{1}{2}(\D-2)(\D-3)\Xi_n(\D-1)
\ee
Therefore $\Xi_n(\D)$ transforms under a representation of the $sl_2(R)_V$ and we can consistently set it to zero without violating the $sl_2(R)_V$ symmetry. Therefore, \eqref{KZ} is indeed $w$ invariant.  

Decoupling of null states gives rise to differential equations which the graviton scattering amplitudes in this theory have to satisfy. So in principle we know a lot about these theories however, solving these equations may not be easy in practice. We hope to return to this in near future. 

\section{Acknowledgements}
We would like to thank Andrew Strominger and Sabrina Pasterski for useful discussion. The work of SB is partially supported by the  Swarnajayanti Fellowship (File No- SB/SJF/2021-22/14) of the Department of Science and Technology and SERB, India and by SERB grant MTR/2019/000937 (Soft-Theorems, S-matrix and Flat-Space Holography). The work of HK is partially supported by the KVPY fellowship of the Department of Science and Technology (DST), Government of India and the Summer Students' Visiting Program at the Institute of Physics, Bhubaneswar, India. The work of PP is supported by an IOE endowed postdoctoral position at IISc, Bengaluru, India.





\end{document}